\documentclass[onecolumn,aps,superscriptaddress]{revtex4}
\usepackage{epsfig,graphics,color}
\newcommand{\be}{\begin{equation}}
\newcommand{\ee}{\end{equation}}
\newcommand{\bea}{\begin{eqnarray}}
\newcommand{\eea}{\end{eqnarray}}

\newcommand{\bc}{\begin{center}}
\newcommand{\ec}{\end{center}}

\newcommand{\forget}[1]{}

\begin{document}

\preprint{}
\title{Latency in local, two-dimensional, fault-tolerant quantum computing}
\author{Federico M. Spedalieri}
\email{federico@ee.ucla.edu}
\affiliation{Department of Electrical Engineering, University
of California, Los Angeles, Los Angeles, California 90095}
\author{Vwani P. Roychowdhury}
\affiliation{Department of Electrical Engineering, University
of California, Los Angeles, Los Angeles, California 90095}

\date{\today}

\begin{abstract}
We analyze the latency of fault-tolerant quantum computing
based on the 9-qubit Bacon-Shor code using a local, two-dimensional
architecture. We embed the data qubits in a 7 by 7 array of
physical qubits, where the extra qubits are used for ancilla preparation
and qubit transportation by means of a SWAP chain. The latency
is reduced with respect to a similar implementation using
Steane's 7-qubit code~\cite{svore2007a}. Furthermore, the
error threshold is also improved to $2.02 \times 10^{-5}$, when
memory errors are taken to be one tenth of the gate error rates. 
\end{abstract}

\maketitle

\section{Introduction}

The fragility of quantum information has always been one
of the most important obstacles for the development
of practical implementations of quantum computing~\cite{nielsen2000}.
The theory of fault-tolerant quantum computation was developed
to deal with the problem of computing in the
presence of errors, even when the quantum gates
required in this process are not perfectly reliable~\cite{preskill1998a}. 
One of the most important achievements in the field, the
threshold theorem, assure us that an arbitrarily
long quantum computation can be performed as long as
the error rates of the faulty gates are below a certain
error threshold~\cite{aharonov1997a}.

The price of dealing with errors and faulty
gates is paid in terms of the space and time overhead 
of the implementation. To reduce the error rates of 
encoded gates, a quantum error-correcting code is 
concatenated with itself (or other code), requiring
more space and time to perform a reliable encoded gate.
On top of this, many physical implementation impose
more constraints on the design of fault-tolerant
circuits. The most important is the locality of
interactions, which forces us to move
qubits next to each other whenever we 
want them to interact through a quantum gate.
We can see that the characteristics of each implementation
will be affected by its underlying architecture. 
Fault-tolerant designs have been studied that 
use different error correcting codes~\cite{steane2003a,gottesman1998a}
and consider different constraints 
and architectures~\cite{szkopek2006a,svore2007a}. 

For many possible implementations of
quantum computing (such as solid state, 
ions in optical lattices, superconducting
qubits) the architecture is not only local
but also restricted in dimensionality. 
In particular, a two-dimensional architecture seems
appealing, since we want to be able to manipulate single
qubits using classical controls, and a planar architecture
will leave us room to do that. This constraint, together
with the locality of interactions will clearly affect
the latency of the computation, since a large amount
of qubit transport will need to be accomplished. It is
then important to study how to optimize this transport
in order to reduce the time overhead. In this paper 
we analyze this problem for a local, two-dimensional
architecture. Our approach is to choose the error correction
code and the implementation of encoded gates in order to
minimize the time required, while at the same time 
trying to keep the space overhead as small as possible.

The paper is organized as follows. In Section \ref{latency}
we begin by discussing the basic ideas of fault-tolerance
and identify the error-correction procedure as the main
contribuitor to the latency of concatenated implementations.
In Section \ref{baconshor} we discuss the properties
of the 9-qubit Bacon-Shor error-correcting code~\cite{shor1995a,bacon2006a}. In
Section \ref{faulttEC} we take a look at the
very useful properties of the fault-tolerant error
correction procedure for the 9-qubit code. In Section
\ref{local} we show how to implement fault-tolerant
encoded gates using this code in a local, two-dimensional
architecture, paying special attention to the latency
of the encoded gates. In Section \ref{errorthreshold}
we compute the error threshold for the Clifford-group
gates. Finally, in Section \ref{conclusions}
we discuss the results and present our conclusions.

\section{Latency in fault-tolerant circuit design}
\label{latency}

Quantum encoding and fault-tolerant circuit design are the two
main tools used in dealing with errors that occur during the
operation of quantum gates in the implementation of a quantum circuit.
By encoding quantum information in the state of several qubits
we can make it more robust to the effects of errors. If these
errors can be detected and corrected, the encoded quantum information
can be preserved. However, since the error detection and 
correction must be implemented with the same faulty gates,
especial care must be put into preventing possible errors from
propagating too much during the computation. Fault-tolerant
design deals exactly with this problem.

While encoding reduces the logical error rate of the computation,
this reduction might not be enough to allow the desired 
computation to be preformed with a high probability of success.
To reduce this error rate even further, the encoding can be
concatenated with itself (or some other encoding). Concatenation
is very powerful in supressing the error rate. If $\epsilon_{phys}$ is
the physical error rate, then the logical error rate $\epsilon_L$
is given by
\be
\label{threshold}
\epsilon_L = \epsilon_0 \left(\frac{\epsilon_{phys}}{\epsilon_0}\right)^{2^k},
\ee
where $k$ is the number of levels of concatenation, and $\epsilon_0$
is the error threshold. We can see that if the physical error rate
is below the error threshold, the logical error rate is suppressed
superexponentially. The error threshold depends on the details of
the encoding and its implementation.

The price of using concatenation to reduce the error rate is
paid as an increase on the size of the circuit, both in
depth and width. This can be seen from the canonical precedure
for implementing a given quantum circuit in a faul-tolerant
way at each level of concatenation. To do this we replace each
physical qubit by an encoded block, and each gate by an encoded
gate. On top of that, after every encoded gate we perform 
error correction on the encoding block. Both the encoded gates
and the error correction procedure must be fault-tolerant
themselves, to prevent errors from propagating to other 
blocks before they can be corrected. This procedure 
generates a self-similar structure as can be seen in
Figure 1.  
\begin{figure}[ht]
\centerline{\includegraphics[scale=0.5]{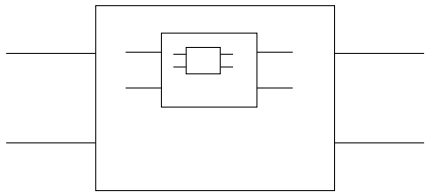}}
\caption{\label{Fig:RecursiveStructure} Recursive structure of concatenated 
implementations of quantum error correction.}
\end{figure}
In this paper we are interested in studying the impact on
latency (or circuit depth) of concatenation, in particular
when other constraints are present. Let $L_{EC}$ be the
latency of the error correction routine (measured in number
of time steps required) applied to
every block. Since error correction is performed
after every encoded gate, if the circuit has $k$ levels
of concatenation, the blowup factor with respect to the 
latency of the unencoded circuit will be \emph{at least}
$L_{EC}^k$. Other contributions to this factor will be given
by the latency of the encoded gates. But for many of the most
used quantum codes, encoded gates can be performed transversally
(requiring only one time step) or require at most a few time steps.
Thus, the latency of the implementation is most heavily influenced
by the latency of the error correcting routine. Finding
fast (in terms of latency) error correcting procedures becomes then 
the key element in trying to reduce the overall latency.

Let us first take a look at the structure of 
error correction. There are some basic steps that any
procedure that corrects errors must follow. First, 
we need to prepare a suitable ancilla state. This 
requires preparing single qubits in a certain states,
such as the eigenstates of one of the Pauli operators.
This will require at least one time step. The ancilla
state will also require entanglement, and hence we need
at least another time step to apply the two-qubit
operations required to produce it (we might need more
than just one time step). Once the ancilla has been prepared,
we need to make it interact with the qubits in the encoding
block. Again, this process takes at least one time step, but
may in general take more (for CSS coded, the data qubits interact
with ancilla qubits twice to detect $X$ and $Z$ errors respectively.)
Finally, the ancilla has to be measured to extract the error syndrome,
which requires another time step. The actual error correction can be
postponed by updating the Pauli frame, as long as the gates 
we apply belong to the Clifford group.

We can see then that performing error correction will
require \emph{at the very least 4 time steps for any 
implementation}. Hence, to improve latency we need to look
for quantum codes that allow fault-tolerant implementations
of error correction to run in as few time steps as possible.
And we have given a sort of benchmark for what ``few time steps''
means. But we are also interested in studying this problem
under other real-world constraints, like locality and 
a two-dimensional architecture. We will show an implementation
with exactly those constraints that can perform error correction
in 7 time steps.

\section{The Bacon-Shor nine-qubit code}
\label{baconshor}

Our implementation will take advantage of some
of the useful properties of the Bacon-Shor 9-qubit code.
We will only give a basic description of this code here, since
a more detailed description can be found 
elsewhere~\cite{shor1995a,bacon2006a}.

The Bacon-Shor 9-qubit code is a stabilizer CSS code that encodes
one logical qubit into nine physical qubits. The distance of this code
is 3, so using the standard notation it is a [[9,1,3]] code. To describe
the properties of this code it will be useful to consider
the nine physical qubits as if they were placed on the vertices of 
a $3\times 3$ lattice as seen in Figure 2. 
\begin{figure}[ht]
\centerline{\includegraphics[scale=0.5]{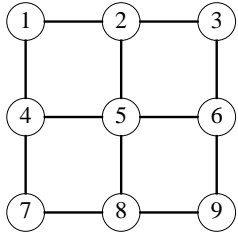}}
\caption{\label{Fig:array} Arrangement of data qubits for
the 9-qubit Bacon-Shor code.}
\end{figure}

The code can be defined by the stabilizer group
${\cal S}$ generated by the stabilizer operators
\bea
\label{stabilizer}
S_1 &=& X_1 X_2 X_3 X_4 X_5 X_6 \nonumber \\
S_2 &=& X_4 X_5 X_6 X_7 X_8 X_9 \nonumber \\
S_3 &=& Z_1 Z_2 Z_4 Z_5 Z_7 Z_8 \nonumber \\
S_4 &=& Z_2 Z_3 Z_5 Z_6 Z_8 Z_9,
\eea
where $X_i$ and $Z_i$ represent the usual Pauli operators applied
to the $i-th$ qubit. Operators $S_1$ and $S_2$ correspond to 
$X$ operators applied to the first and second row of qubits, and the
second and third row, respectively. In a similar manner, $S_3$ and $S_4$
correspond the $Z$ operators applied to the first and second column of
qubits, and the second and third column, respectively.

The different syndromes (i.e., the vector of eigenvalues of the 
four opeators in (\ref{stabilizer})) induce a decomposition
of the Hilbert space of the nine qubits in the code block into
subspaces encoding actually five logical qubits. In each one
of these subspaces we can define a subsystem decomposition 
and write
\be
\label{subsystemdec}
{\cal H} = \bigoplus_{syndromes} ({\cal H}_L \otimes {\cal H}_T),
\ee
where the direct sum is over all possible syndromes, ${\cal H}_L$ is
the Hilbert space of the logical qubit that will be fully protected
by the code, and ${\cal H}_T$ is the Hilbert space of the remaining 
encoded qubits (which will not be fully protected).

The logical operators associated with the encoded qubit are
given by 
\bea
\label{logicalops}
X_L &=& X_1 X_2 X_3 \nonumber \\
Z_L &=& Z_1 Z_4 Z_7.
\eea
We can see that $X_L$ corresponds to $X$ operators
acting on all the qubits on the first row of Figure \ref{Fig:array}
while $Z_L$ corresponds to $Z$ operators acting on the first column.
It si easy to check that these logical operators commute with
the stabilizer generators. The logical operators for qubits
encoded in ${\cal H}_T$ can be chosen from the nonabelian group
\bea
\label{Tgroup}
T&=&\langle X_1 X_4,X_4 X_7,X_2 X_5,X_5 X_8,X_3 X_6,X_6 X_9, \nonumber \\
 & & \ \ Z_1 Z_2,Z_2 Z_3,Z_4 Z_5,Z_5 Z_6,Z_7 Z_8,Z_8 Z_9\rangle.
\eea    
It is easy to see that any operator in $T$ commutes not only with
the stabilizer generators, but also with the logical operators
in (\ref{logicalops}). This leads to one of the key features of
this code (and of subsystem eoncoding in general): the state
of the logical qubit is not uniquely encoded. Once this
state has been encoded in the nine qubit block, applying
any operator in $T$ does not affect the encoded quantum
information, since these operators commute with the logical
operators $X_L$ and $Z_L$. This simplifies error correction
and makes some two-qubit errors actually trivial.
 
\section{Faul-tolerant error correction}
\label{faulttEC}

Since the 9-qubit Bacon-Shor code is a CSS code, we can 
apply the techniques introduced by Steane to perform 
fault-tolerant error correction. This requires preparing 
the logical state $|0\rangle$, a simultaneous eigenstate of stabilizer
generators and the logical
operator $Z_L$ with eigenvalues +1. As we pointed out before, this does
not completely determine the encoded state, since we can apply any operator
in the set $T$ without perturbing the encoded information. This ambiguity
can be fixed by requiring the encoded state to be an eigenstate
of the operators $X_1 X_4,X_4 X_7,X_2 X_5,X_5 X_8,X_3 X_6,X_6 X_9$ 
(which belong to $T$) with eigenvalue +1. It is not difficult to
check that the only state satisfying these constraints is given by
\be
\label{logical0}
|\bar{0}\rangle = \frac{1}{\sqrt{8}}(|+++\rangle_{147} +|---\rangle_{147})
(|+++\rangle_{258} +|---\rangle_{258})(|+++\rangle_{369} +|---\rangle_{369}),
\ee
where $|\pm\rangle = \frac{1}{\sqrt{2}}(|0\rangle \pm |1\rangle)$, as usual.
Geometrically, this state corresponds to the tensor product of three
``cat'' states in the Hadamard rotated basis, each one comprising
the qubits in each column in Figure \ref{Fig:array}. Similarly, it is not 
difficult to see that the encoded $|\bar{+}\rangle$ state corresponds
to three ``cat'' states in the computational basis, each one now 
lying across each one of the three rows of qubits in Figure \ref{Fig:array}.
This geometric description will be very useful when we look at 
how to prepare and use these states in a local, two-dimensional architecture.

\begin{figure}[ht]
\centerline{\includegraphics[scale=0.5]{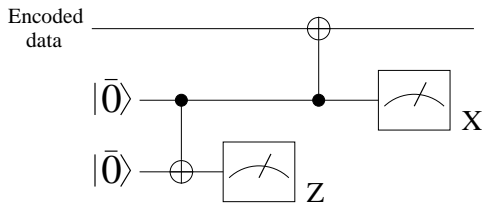}}
\caption{\label{Fig:SteaneSyndrome} Quantum circuit for Steane's 
fault-tolerant syndrome extraction.}
\end{figure}

To extract the error syndrome in the Steane fault-tolerant error
correction scheme, we need to apply the circuit in Figure \ref{Fig:SteaneSyndrome}.
To look for $Z$ errors on the date block, an ancilla block encoded in the
$|\bar{0}\rangle$ state interacts with the data block through
an encoded CNOT gate (which for this code can be implemented transversally.)
The qubits in the ancilla block are then measured in the $X$ basis,
and the results of these measurements are used to classically compute
the error syndrome. But before the ancilla block interacts with the data,
it needs to be verified using an extra ancilla block also encoded
in the $|\bar{0}\rangle$ state, to check for possible $X$ errors
in the ancilla block that could propagate to the data. To make the procedure
fault-tolerant, this verification step needs to be carried out three times,
and the ancilla block is accepted if and only if no more than one verification
step failed. A similar scheme is used to check the data qubits for
$X$ errors.

The ancilla verification state is the main contributor to the
latency of the error correction procedure, since we need
to construct a verification block at least two times before
allowing the ancilla block to interact with the data. 
Fortunately, Aliferis and Cross~\cite{aliferis2007a} showed that for the
9-qubit code using Steane's procedure  \emph{the verification
step is not required}. This property can be somewhat traced back
to the fact that the encoded states $|\bar{0}\rangle$ and $|\bar{+}\rangle$
break down into three entangled ``cat'' states of only three qubits each, and these
``cat'' states are also rather robust against pairs of
errors that could be introduced by a faulty gate during their preparation.
This decomposition into a tensor product of states with less qubits
also makes the preparation process efficient in terms of latency.  

\begin{figure}[ht]
\centerline{\includegraphics[scale=0.5]{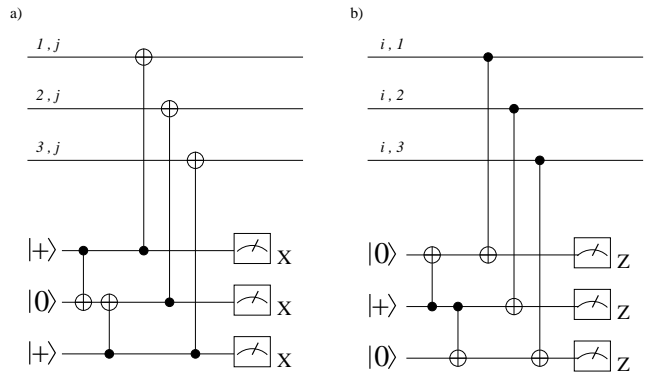}}
\caption{\label{Fig:ACSyndrome} Fault-tolerant circuit for syndrome
extraction for the 9-qubit Bacon-Shor code, introduced
by Aliferis and Cross.}
\end{figure}

The circuit for syndrome extraction takes then a very simple 
form, as can be seen in Figure \ref{Fig:ACSyndrome}.
From this figure we can see that the latency for the error
correction routine is 6 time steps. Preparing the ancilla
requires 3 time steps: one for single-qubit preparations and
two to entangle them to form a ``cat'' state (or its Hadamard
rotated version). The interaction with the data block through
a transversal CNOT can be accomplished in one time step and
so can the final ancilla measurements. Even though each
syndrome extraction can be accomplished in 5 time steps,
we need to delay one of them by 1 time step since the
two ancilla blocks cannot interact with the data at the same time,
and this gives us a total latency of 6 time steps. And as we discussed
before, instead of actually correcting an error we will only update the
Pauli frame. This can be done as long as we apply gates that belong
to the Clifford group, since they preserve the tensor product form of
the errors. We will see that we can relegate non-Clifford gates
to the higest level of concatenation (i.e., the algorithmic level),
so we can run all the lower levels without physically performing
Pauli operations to correct the errors.

\section{Local, two-dimensional architecture}
\label{local}

The fault-tolerant error correction procedure described above
relies on an abstract architecture in which there is no restriction
on the interaction between qubits. A two-qubit gate
can be applied to any pair pair of qubits in one time step.
In reality, a physical implementation of quantum computing
will usually impose a locality constraint that requires that two interacting
qubits are actually next to each other. Then, to implement an
abstract quantum circuit we will need to introduce an information
transport process that is based on local interactions. The simplest
way of doing that is to use a chain of SWAP gates, in which the
states of two neighboring qubits is exchanged. However, we need to be
careful if we want to preserve the fault-tolerance of the the scheme,
since a faulty SWAP might introduce two errors in one block if it is applied
to two data qubits. 

Besides the locality of the interaction, as a matter of practical design
it is advantageous to lay our qubits in some kind of two-dimensional
array. This would allow the unavoidable classical control signals
to have easier access to each individual qubit in order to perform
necessary operations like single qubit preparation and measurement, and
single qubit gates. Then the problem becomes how to design a fault-tolerant
error correction procedure that minimizes the impact on latency
of the locality constraint on a two-dimensional architecture, trying to
keep the space overhead under control and without lowering too much the
error threshold. 

In~\cite{svore2007a}, Svore, DiVincenzo and Terhal studied this
type of architecture, although their main objective was to compute the
error threshold for that implementation. Also they used Steane's 
7-qubit code to encode the quantum information and applied
Steane's scheme for error correction. They embedded the seven data
qubits in a $6 \times 8$ array of physical qubits. The extra qubits
where used for ancilla preparation and measurement, and for qubit
transport by means of SWAP gates. We will consider a similar type
of implementation based on the 9-qubit Bacon-Shor code.

\begin{figure}[ht]
\centerline{\includegraphics[scale=0.8]{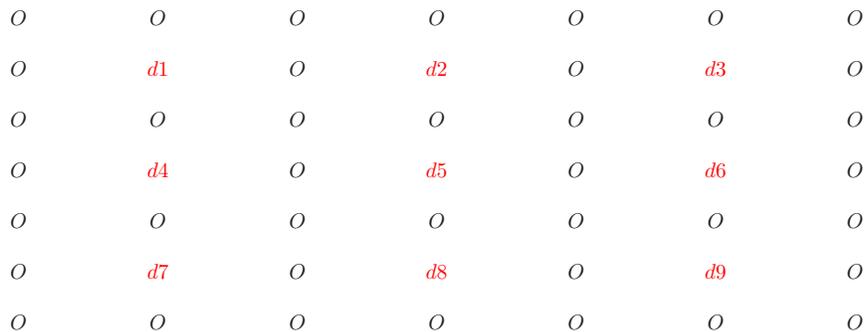}}
\caption{\label{Fig:7x7array} Encoding block: the data qubits are embeded
in a 7 by 7 array of physcical qubits. The remaining ``dummy'' qubits
(noted by the letter $O$) are used for ancilla preparation and
qubit transportation.}
\end{figure}

Our implementation consists of embedding the nine data qubits
of the 9-qubit code in a $7 \times 7$ array of physical qubits.
During error correction, the data qubits are located at the
positions shown in Figure \ref{Fig:7x7array}.
The remaining qubits (represented by $O$) act both as the
workspace in which ancillas are prepared and measured, and as
the ``rails'' on which data and ancilla qubits are transported
using chains of SWAP gates.      

\subsection{Error Correction}

As it can be seen in Figure \ref{Fig:ACSyndrome}, to perform error correction we need
to prepare three ``cat'' states, each interacting with   
one (and only one) of the rows of data qubits in Figure \ref{Fig:7x7array}, and
three Hadamard-rotated ``cat'' states to interact with the three 
columns of data qubits. For the first and last row (column),
we can use the external rows (columns) in Figure  \ref{Fig:7x7array}, to prepare
the required ancilla and move some of these ancilla qubits 
next to the corresponding data qubit. Each of these parts
of the error correction can be performed idependently of 
one another. But for the ancilla states that interact with 
middle row and column, we need to carefully time their 
preparation, movement and interaction with the data, because
they make use of the same workspace. This additional
complication (due to the locality and two-dimensional
character of this implementation) increases the latency
of the error correction procedure with respect to
the abstract case by only one time step. 

\begin{figure}[ht]
\centerline{\includegraphics[scale=0.8]{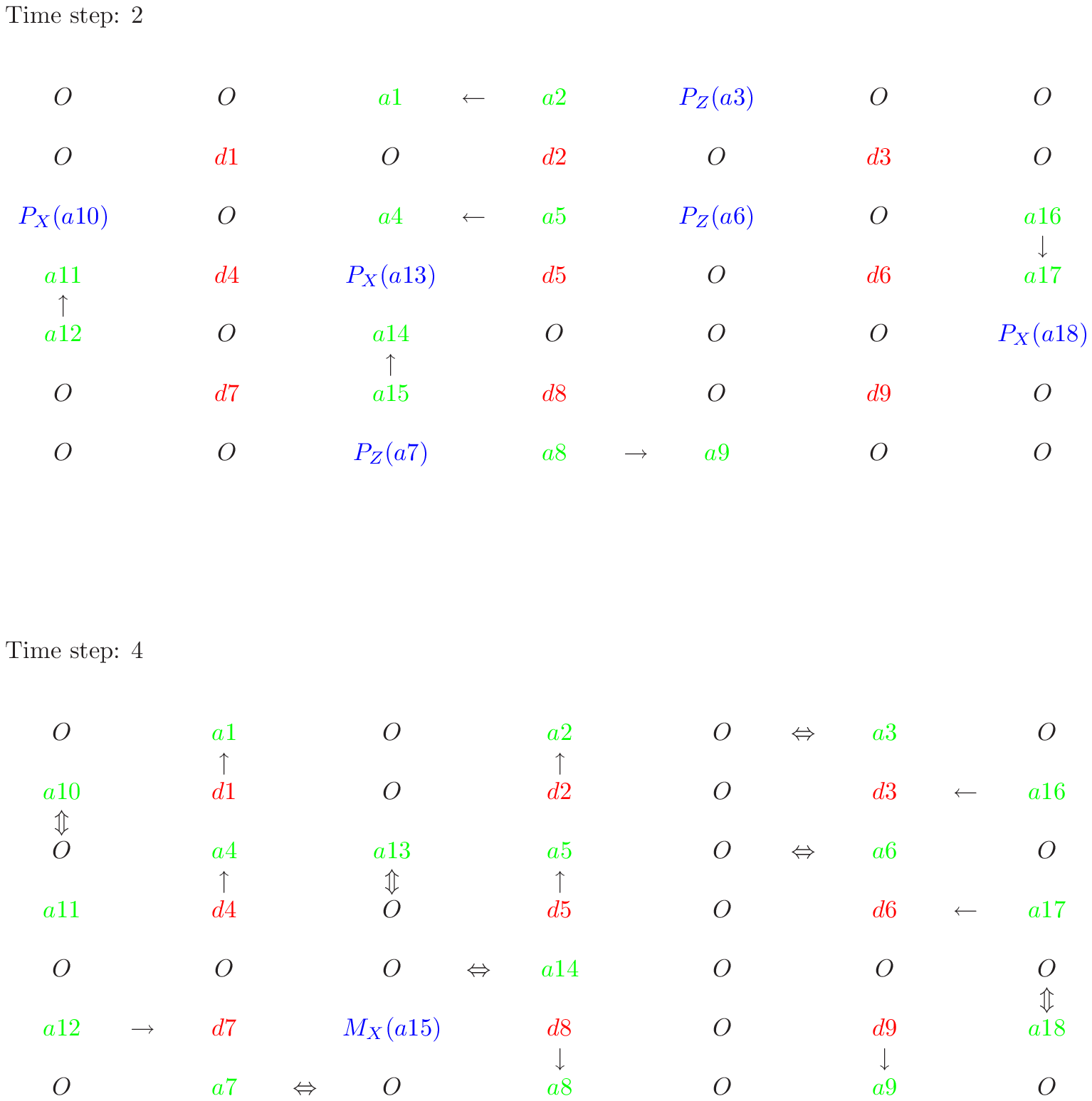}}
\caption{\label{Fig:ECsnapshot} Two snapshots of the error
correction procedure.}
\end{figure}

In Figure \ref{Fig:ECsnapshot}, we present two snapshots of the error correction
procedure (the complete sequence is presented in the appendix). 
The data qubits are denoted by $d1$, $d2$, etc., 
and remain fixed during the error correction procedure.
The ancilla qubits (denoted by $a1$, $a2$, etc.) are prepared
and transported using the workspace of dummy qubits (denoted by 
$O$.) The preparation of ancilla qubits is noted by $P_{X,Z}(a_j)$,
and represents the preparation of the qubit in the eigenstate 
of either $X$ or $Z$, with eigenvalue 1. Note that not all ancilla
qubits are prepared at the same time, in order to reduce
the number of memory error locations. The single arrows represent
CNOT gates, pointing from control to target qubits. The double-headed
arrows represent SWAP gates. In time step 2
we can see the second step in the preparation of all the required
cat states. In time step 4 we can see the data qubits interacting 
with the ancilla, some ancilla qubits still moving via SWAP gates
to get in position, and one ancilla qubit being measured (noted
by $M_{X,Z}()$.)

\subsection{Encoded gates}

For the purpose of concatenation we need to be able to
perform both the encoded CNOT and the encoded SWAP gate.
The encoded SWAP gate is very simple, since it only involves 
moving the data qubits from one encoding block to another.
This is done using SWAP gates. It is clear that the latency
of the encoded SWAP gate is only 7 time steps.

For the 9-qubit code, the CNOT gate is transversal. Then
we only need to move the data qubits so that the corresponding
qubits of two neighboring blocks are next to each
other, and then apply a single-qubit CNOT between each pair. 
The most efficient way of doing this, from the point of view
of latency, is to  interleave the rows (or columns) of the two
blocks. To do this, we first move all the data qubits in one
block up one row (or left one column), while the data qubits on the other block start
moving laterally (or vertically) towards the first block. Then we keep
moving tha data qubits towards each other, interleaving the 
rows (columns) until the corresponding data qubits are next to each
other. Thn we apply a single-qubit CNOT between each pair, and
move all data qubits back to their original positions. In this way,
that latency of the encoded CNOT is only 9 time steps. 

The encoded Hadamard gate can also be easily implemented. As noted 
in~\cite{aliferis2007a}, the encoded Hadamard is equivalent to applying
single-qubit Hadamards to all physical qubits and rotating the $3\times 3$ array
of qubits by 90 degrees. In our implementation, this rotation can be accomplished by
transporting the data qubits in the perimeter of the array in Figure \ref{Fig:7x7array}
(i.e., all qubits except $d5$) along that perimeter until the rows are transformed into
the columns of the array. This transport can be done in 4 time steps, since we
can move all qubits at the same time. Then, the total latency of the encoded Hadamard
gate is only 5 time steps: one to apply single-qubit Hadamards and four to
rotate the array.

\subsection{Encoded preparation and measurement}

Finally, we need to be able to fault-tolerantly prepare 
the encoded $|\bar{0}\rangle$ and $|\bar{+}\rangle$ states, and 
fault-tolerantly measure the encoded operators $X$ and $Z$. 
Since the 9-qubit code is
a CSS code, the fault-tolerant measurement reduces to measuring the
single-qubit operator X (or Z) for each data qubit in the encoding 
block and classically post-processing the outcomes (as is usual, 
we assume that classical processing can be done flawlessly.)

\begin{figure}[ht]
\centerline{\includegraphics[scale=0.5]{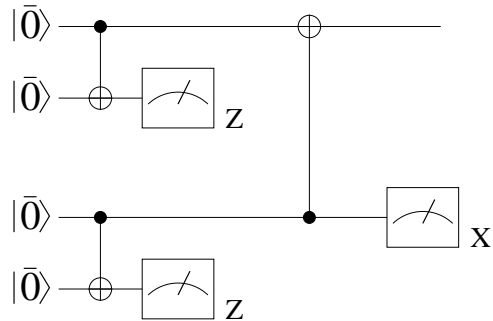}}
\caption{\label{Fig:SteanePrep} Fault-tolerant 
preparation of the encoded $|\bar{0}\rangle$ state.}
\end{figure}

For the fault-tolerant preparation of encoded states, we will use 
a procedure that applies to any CSS code,
and is illustrated in Figure \ref{Fig:SteanePrep} for the encoded state $|\bar{0}\rangle$.
This scheme requires that we prepare four copies of the encoded state.
As discussed above, the state $|\bar{0}\rangle$ is equivalent to 
three ``cat'' states in the Hadamard-rotated basis, each comprised
of the qubits in one of the columns of data qubits in Figure \ref{Fig:7x7array}.
To prepare the four copies we need to use 36 physical
qubits. We can choose them to be the $6 \times 6$ array obtained when we
subtract the first row and column from the $7 \times 7$ array.
We can use 3 rows and 6 columns to prepare two copies
of $|\bar{0}\rangle$, \emph{with their columns interleaved} (giving
us the 4 copies needed in the $6 \times 6$ array.) Once these copies are
prepared we can apply encoded CNOT gates on each pair of
encoded states, which takes only one time step. Then we measure two 
of the copies, and then move the qubits in the columns vertically
using SWAP gates in order to position them next to the corresponding 
qubit in the remaining copy. We then apply another CNOT gate and measure 
one of the copies. The remaining qubits encode the state $|\bar{0}\rangle$.
This preparation procedure requires 9 time steps. A step-by-step pictorial
representation of this preparation can be found in the appendix.
The encoded state $|\bar{+}\rangle$ can be prepared in a similar way,
interleaving rows instead of columns.

\subsection{1-Recs and latency}

A 1-Rectangle (1-Rec) is formed by an encoded gate followed by error correction
on all the encoded blocks involved in the gate (with the exception
of the 1-measurement, that requires only classical post-processing.) 
We can summarize the above results in the following table 
showing the latency in time steps, of each 1-Rec, for both
the 9-qubit code discussed here and the 7-qubit code presented
in~\cite{svore2007a}.
\begin{table}
\label{latencytable}
\begin{tabular}{|l|c|c|}
Gate & 9-qubit & 7-qubit \\ \hline\hline
CNOT & 16 & 35 \\ \hline
SWAP & 14 & 35 \\ \hline
$|0\rangle$-preparation & 16 & 41 \\ \hline
$|+\rangle$-preparation & 16 & 41 \\ \hline
Measurement & 1 & 1 \\ \hline
 \end{tabular} 
\caption{Latencies (in number of time steps) of encoded
gates for the 9-qubit and 7-qubit codes.} 
\end{table}
The longest 1-Rec is the one corresponding to the 
preparation of encoded states. However, since these preparations
are always performed on blocks representing dummy qubits \emph{that are free 
during the previous time step}, we can always start these preparations 
ahead of time, reducing their effective latency. Thus, the lower
limit for speed in this implementation is given by the CNOT 1-Rec,
that requires 16 time steps. This produces a reduction in latency
with respect to the 7-qubit code implementaion in~\cite{svore2007a} 
by a factor of $(16/35)^k \sim 0.45^k$, with $k$ the 
number of concatenation levels.

\subsection{Universality}

Universality of this scheme can be achieved by supplementing the
Clifford-group gates implemented with non-Clifford gates, such
as the phase gate $S$ and its square root the $T$ gate. Since the 9-qubit
code does not allow for a simple, fault-tolerant implementation
of these gates, we need to use a different approach. One possibility,
following~\cite{aliferis2007a} is to purify noisy encoded $|+i\rangle$ 
states using only Clifford-group gates (which are virtually error free at
high concatenation levels), and later use the purified
encoded $|+i\rangle$ state to apply the $S$ gate with higher reliability. 
This can be done at any concatenation level, using the injection by teleportation 
technique developed by Knill~\cite{knill2005a}
to insert a noisy encoded $|+i\rangle$ state at the
required level. This approach is also used in~\cite{svore2007a} to improve
the impact on the threshold of preparing the ancillas required to implement
the $S$ and $T$ gates. There it is shown that the threshold for the
preparation of these ancilla states is higher than the threshold
for Clifford gates. That analysis also applies to the 9-qubit code 
(the only difference being the number of gates involved in the decoding
during the injection by teleportation, and that number is comparable for both codes.)
Once we are able to apply $S$ we can use it
to generate the ancilla state required to apply $T$. These
preparations can be done offline, since these gates are
only required at the algorithmic level.

\section{Error threshold}
\label{errorthreshold}

We compute the error threshold for the case of stochastic, adversarial
noise. Following the procedure presented in~\cite{aliferis2006a} and also
used in~\cite{svore2007a}, we compute the number of malignant pairs of
locations for the CNOT extended rectangle. There are 7 different
types of locations: (1) $|+\rangle$ preparation, (2) $|0\rangle$ preparation,
(3) memory, (4) SWAP gate, (5) $Z$ measurement, (6) $X$ measurement, and
(7) CNOT gate. The error probability for the $k^{th}$ level CNOT extended
rectangle is given by
\be
\label{errork}
\epsilon^{(k)}_{7} \leq \sum_{i\leq j}^7 \alpha_{ij} \epsilon^{(k-1)}_i
\epsilon^{(k-1)}_j  + B (\epsilon^{(k-1)}_{max})^3,
\ee
where $\epsilon^{(l)}_i$ is the error rate of a gate of type $i$ at level $l$,
and $\epsilon^{(k-1)}_{max}$ is the maximum of all the error rates 
(which in our
case is $\epsilon_7^{(k-1)}$.) The second term on the RHS of (\ref{errork}) bounds
the contribution of third and higher order terms, with 
$B=\left(\begin{array}{c} 791 \\ 3 \end{array}\right)$, where 791 is
the number of locations in the CNOT extended rectangle. The coefficients
$\alpha_{ij}$ represent the number of malignant pairs of locations
of each kind, and are given by
\be
\label{malignantmatrix}
\alpha = \left(\begin{array}{ccccccc}
114 & & & & & &  \\
0 & 160 & & & & &  \\
1112 & 1362 & 3027 & & & &  \\
2302 & 2483 & 11160 & 11040 & & &  \\
0 & 402 & 3132 & 1422 & 216 & &  \\
300 & 0 & 3126 & 1416 & 0 & 216 &  \\
1186 & 1501 & 7962 & 14626 & 1740 & 1410 & 4465  \\
\end{array} \right).
\ee
If we take all error rates to be equal, the total number of 
malignant pairs is $A=75,880$. Going back to Eq. (\ref{errork}) we can
write 
\be
\epsilon^{(k)}_{7} \leq A' \left(\epsilon^{(k-1)}_7\right)^2,
\ee 
with $A' = \frac{A}{2}\left(1+\sqrt{1+\frac{4 B}{A^2}}\right)=76,948$.
This gives a lower bound for the accuracy threshold of 
$1.3 \times 10^{-5}$. If we assume (as it is usually done) that 
the memory errors are one tenth of the other errors, the 
threshold becomes $2.02 \times 10^{-5}$. 

\section{Conclusions and discussion}
\label{conclusions}

In this paper we have analyzed the implementation of
fault-tolerant quantum computing in a local, 2-dimensional
setting. In particular, we studied the effects on the latency
of the computation that such constraints will have. Locality
requires that any pair of interacting qubits have
to be placed next to each other before the interaction can
take place. This is accomplished by applying a chain of
SWAP gates, between a qubit and one of its neighbors, until
the two qubits are side by side and the required local logical
gate can be applied. This clearly affects the latency of the 
computation. The fact that this movement
is restricted to a 2-dimensional setting will also impact
the latency when several moving qubits have to be maneuvered
to avoid each other.

Another issue is that
our best approach to implement fault-tolerant quantum computing,
through concatenation of error-correcting codes, yields an
exponential increase in latency with the number of concatenation
levels. This is not as bad as it may seem, since we would only
need a finite number of concatenation levels. But from a
practical point of view it will translate into huge latency overheads.

The answer to minimizing the latency increase due to concatenation
rests on a clever choice of the error-correcting code. Since
concatenation requires the application of error correction after
every single encoded gate at every encoded level, it is clear that an
encoding with 
a fast (in terms of latency) error-correction procedure is 
highly desirable. And because we are restricted to local
operations, and this need of
transporting qubits is at the heart of the latency increase,
it is reasonable to look at small codes that can be embeded 
in a compact 2-dimensional architecture that minimizes qubit
transport. This is the setting in which the 9-qubit code
has shown to posses remarkable properties that work to our
advantage.

Furthermore, a simple analysis of error correction shows
that the latency of this process has a lower bound. 
To perform error correction we need to prepare and
entangle ancilla qubits, make them interact with the data and
measure them. All these steps are unavoidable and translate
into a lower bound for the latency of the concatenated circuit.
The 9-qubit code property of not requiring ancilla verification
is the origin of its short latency. It is probably as
good as we can get it to be. And implementing it
in a local, 2-dimensional architecture only increases the
latency from 6 to 7 time steps. This results in a very 
fast and compact implementation of the encoded CNOT 1-Rec.
Compared to a similar implementation~\cite{svore2007a} that uses 
Steane's 7-qubit code, the 9-qubit code improves the latency
of the CNOT 1-Rec (encoded gate plus error correction) by a 
factor of $0.45$, which translates into an improvement in overall
latency of $(0.45)^k$, with $k$ the number of concatenation levels.

The 9-qubit code implementation also improves the
accuracy threshold for Clifford-group gates (the only
ones needed for error correction.) If we take all errors
to have the same rate, the thresold becomes $1.3 \times 10^{-5}$,
compared to $1.1 \times 10^{-5}$ for the 7-qubit code.
If we take memory errors to have a lower error rate (one tenth
of other error rates), we obtain a threshold of
$2.02 \times 10^{-5}$ ($1.85 \times 10^{-5}$ for the 7-qubit code.)
The threshold increase is modest, but the latency increase is
sizable. Furthermore, the space overhead is essentially the same
($49$ qubits in the encoding block for the 9-qubit code versus
$48$ qubits for the 7-qubit code), but the 9-qubit implementation
has the added feature of treating encoded gates the same way
whether the encoded blocks are horizontally adjacent or
vertically adjacent, as opposed to the 7-qubit code whose
$6$ by $8$ basic array of physical qubits necessarily needs
to treat these two operations differently.

Universality can be achieved by supplementing the
Clifford-group gates implemented, with non-Clifford gates such
as the phase gate $S$ and its square root the $T$ gate. Since the 9-qubit
code does not allow for a simple, fault-tolerant implementation
of these gates, this is accomplished by the preparation
of special ancilla states that can be used to apply the required gates.
Even though the preparation of these ancilla states is not fault-tolerant,
we can combine the techniques of injection by teleportation and distillation
to prepare the required states with high fidelity at the desired
level of concatentation. These
preparations can be done offline, since these gates are
only required at the algorithmic level. How to 
integrate these preparations into a 2-dimensional architecture
in an efficient manner is the next step that needs to be addressed in this
implementation.

\appendix

\section{Detailed implementation of error correction and encoded state preparation}

Here we present a pictorial representation of each step required to perform
both error correction and encoded state preparations.
We have already shown a couple of snapshots of the error correction procedure
in Figure \ref{Fig:ECsnapshot}. Here we present the complete sequence. 
Data qubits are represented by $d1,\ d2,$ etc., and ancilla qubits are noted
as $a1,\ a2,$ etc. The dummy qubits used to prepare and mesaure the ancilla as well
as for qubit transport, are represented by $O$. The preparation of ancilla qubits
is noted by $P_{X,Z}(a_j)$, which represents the preparation of the qubit in the 
eigenstate of either $X$ or $Z$, with eigenvalue 1. The single-headed arrows represent
CNOT gates, pointing from control to target qubits; the double-headed arrows represent
SWAP gates; and qubit measurements in the $X$ or $Z$ bases are represented by $M_{X,Z}()$.

\subsection{Error correction}

\vbox{

\small

Time step: 0
\\

$
\begin{array}{ccccccccccccc}
\textcolor{white}{P_Z(a11)}& \textcolor{white}{\Leftrightarrow} &\textcolor{white}{P_Z(a11)} & \textcolor{white}{\Leftrightarrow} &\textcolor{white}{P_Z(a11)} & \textcolor{white}{\Leftrightarrow} & \textcolor{white}{P_Z(a11)} &\textcolor{white}{\Leftrightarrow} &\textcolor{white}{P_Z(a11)} &\textcolor{white}{\Leftrightarrow} & \textcolor{white}{P_Z(a11)} & \textcolor{white}{\Leftrightarrow} & \textcolor{white}{P_Z(a11)} \\

O  & & O & & O  & & O  & & O & & O  & & O  \\
  & &   & &   & &    & &   & &  & &   \\
O & & \textcolor{red}{d1} & & O & & \textcolor{red}{d2} & & O & & \textcolor{red}{d3} & & O \\ 
  & &   & &  & &    & &   & &   & &   \\
O & & O &  & O & & O & & O & & O & & O  \\
  & &  & &   & &    & &   & &   & &    \\
O & & \textcolor{red}{d4} & & O & & \textcolor{red}{d5} & & O & & \textcolor{red}{d6} & & O \\
  & &   & &   & &    & &   & &   & &     \\
O & & O & & O & & O & & O & & O & & O \\
  & &   & &   & &    & &   & &   & &    \\
O & & \textcolor{red}{d7} & & O & & \textcolor{red}{d8} & & O & & \textcolor{red}{d9} & & O \\
   & &  & &   & &    & &   & &   & &    \\
O & & O &  & O & & O & & O & & O & & O \\
   & &   & &   & &    & &   & &   & &  

\end{array}
$
}

\vbox{

\small

Time step: 1
\\

$
\begin{array}{ccccccccccccc}
\textcolor{white}{P_Z(a11)}& \textcolor{white}{\Leftrightarrow} &\textcolor{white}{P_Z(a11)} & \textcolor{white}{\Leftrightarrow} &\textcolor{white}{P_Z(a11)} & \textcolor{white}{\Leftrightarrow} & \textcolor{white}{P_Z(a11)} &\textcolor{white}{\Leftrightarrow} &\textcolor{white}{P_Z(a11)} &\textcolor{white}{\Leftrightarrow} & \textcolor{white}{P_Z(a11)} & \textcolor{white}{\Leftrightarrow} & \textcolor{white}{P_Z(a11)} \\

O  & & O  & & \textcolor{blue}{P_Z(a1)}  & & \textcolor{blue}{P_X(a2)}  & & O  & & O  & & O  \\
  & &   & &   & &    & &   & &   & &   \\
O & & \textcolor{red}{d1} & & O & & \textcolor{red}{d2} & & O & & \textcolor{red}{d3} & & O \\ 
  & &   & &   & &    & &   & &   & &   \\
O & & O & & \textcolor{blue}{P_Z(a4)} & & \textcolor{blue}{P_X(a5)} & & O & & O & & \textcolor{blue}{P_X(a16)} \\
   & &   & &   & &    & &   & &   & &    \\
\textcolor{blue}{P_Z(a11)} & & \textcolor{red}{d4} & & O & & \textcolor{red}{d5} & & O & & \textcolor{red}{d6} & & \textcolor{blue}{P_Z(a17)} \\
   & &   & &   & &    & &   & &   & &    \\
\textcolor{blue}{P_X(a12)} & & O & & \textcolor{blue}{P_Z(a14)} & & O & & O & & O & & O \\
   & &   & &   & &    & &   & &   & &    \\
O & & \textcolor{red}{d7} & & \textcolor{blue}{P_X(a15)} & & \textcolor{red}{d8} & & O & & \textcolor{red}{d9} & & O \\
   & &   & &   & &    & &   & &   & &    \\
O & & O & & O & & \textcolor{blue}{P_X(a8)} & & \textcolor{blue}{P_Z(a9)} & & O & & O \\
   & &   & &   & &    & &   & &   & &  

\end{array}
$
}

\vbox{

\small

Time step: 2
\\

$
\begin{array}{ccccccccccccc}
\textcolor{white}{P_Z(a11)}& \textcolor{white}{\Leftrightarrow} &\textcolor{white}{P_Z(a11)} & \textcolor{white}{\Leftrightarrow} &\textcolor{white}{P_Z(a11)} & \textcolor{white}{\Leftrightarrow} & \textcolor{white}{P_Z(a11)} &\textcolor{white}{\Leftrightarrow} &\textcolor{white}{P_Z(a11)} &\textcolor{white}{\Leftrightarrow} & \textcolor{white}{P_Z(a11)} & \textcolor{white}{\Leftrightarrow} & \textcolor{white}{P_Z(a11)} \\

O  & & O  & & \textcolor{green}{a1}  & \leftarrow & \textcolor{green}{a2}  & & \textcolor{blue}{P_Z(a3)}  & & O  & & O  \\
  & &   & &   & &    & &   & &   & &   \\
O & & \textcolor{red}{d1} & & O & & \textcolor{red}{d2} & & O & & \textcolor{red}{d3} & & O \\ 
  & &   & &   & &    & &   & &   & &   \\
\textcolor{blue}{P_X(a10)} & & O & & \textcolor{green}{a4} & \leftarrow & \textcolor{green}{a5} & & \textcolor{blue}{P_Z(a6)} & & O & & \textcolor{green}{a16} \\
  & &   & &   & &    & &   & &   & & \downarrow   \\
\textcolor{green}{a11} & & \textcolor{red}{d4} & & \textcolor{blue}{P_X(a13)} & & \textcolor{red}{d5} & & O & & \textcolor{red}{d6} & & \textcolor{green}{a17} \\
\uparrow   & &   & &   & &    & &   & &   & &    \\
\textcolor{green}{a12} & & O & & \textcolor{green}{a14} & & O & & O & & O & & \textcolor{blue}{P_X(a18)} \\
   & &   & & \uparrow  & &    & &   & &   & &    \\
O & & \textcolor{red}{d7} & & \textcolor{green}{a15} & & \textcolor{red}{d8} & & O & & \textcolor{red}{d9} & & O \\
   & &   & &   & &    & &   & &   & &    \\
O & & O & & \textcolor{blue}{P_Z(a7)} &  & \textcolor{green}{a8} & \rightarrow & \textcolor{green}{a9} & & O & & O \\
   & &   & &   & &    & &   & &   & &  

\end{array}
$

}

\vbox{

\small

Time step: 3
\\

$
\begin{array}{ccccccccccccc}
\textcolor{white}{P_Z(a11)}& \textcolor{white}{\Leftrightarrow} &\textcolor{white}{P_Z(a11)} & \textcolor{white}{\Leftrightarrow} &\textcolor{white}{P_Z(a11)} & \textcolor{white}{\Leftrightarrow} & \textcolor{white}{P_Z(a11)} &\textcolor{white}{\Leftrightarrow} &\textcolor{white}{P_Z(a11)} &\textcolor{white}{\Leftrightarrow} & \textcolor{white}{P_Z(a11)} & \textcolor{white}{\Leftrightarrow} & \textcolor{white}{P_Z(a11)} \\

O  & & \textcolor{green}{a1} & \Leftrightarrow & O  & & \textcolor{green}{a2}  & \rightarrow & \textcolor{green}{a3}  & & O  & & O  \\
  & &   & &   & &    & &   & &   & &   \\
O & & \textcolor{red}{d1} & & O & & \textcolor{red}{d2} & & O & & \textcolor{red}{d3} & & \textcolor{green}{a16} \\ 
  & &   & &   & &    & &   & &   & & \Updownarrow  \\
\textcolor{green}{a10} & & \textcolor{green}{a4} & \Leftrightarrow & O & &\textcolor{green}{a5} & \rightarrow & \textcolor{green}{a6} & & O & & O  \\
\downarrow  & &   & &   & &    & &   & &   & &    \\
\textcolor{green}{a11} & & \textcolor{red}{d4} & & \textcolor{green}{a13} & & \textcolor{red}{d5} & & O & & \textcolor{red}{d6} & & \textcolor{green}{a17} \\
  & &   & & \downarrow  & &    & &   & &   & & \uparrow    \\
O & & O & & \textcolor{green}{a14} & & O & & O & & O & & \textcolor{green}{a18} \\
\Updownarrow   & &   & &   & &    & &   & &   & &    \\
\textcolor{green}{a12} & & \textcolor{red}{d7} & & \textcolor{green}{a15} & \rightarrow & \textcolor{red}{d8} & & O & & \textcolor{red}{d9} & & O \\
   & &   & &   & &    & &   & &   & &    \\
O & & O & & \textcolor{green}{a7} & \leftarrow & \textcolor{green}{a8} & & O & \Leftrightarrow & \textcolor{green}{a9} & & O \\
   & &   & &   & &    & &   & &   & &  

\end{array}
$

}

\vbox{

\small

Time step: 4
\\

$
\begin{array}{ccccccccccccc}
\textcolor{white}{P_Z(a11)}& \textcolor{white}{\Leftrightarrow} &\textcolor{white}{P_Z(a11)} & \textcolor{white}{\Leftrightarrow} &\textcolor{white}{P_Z(a11)} & \textcolor{white}{\Leftrightarrow} & \textcolor{white}{P_Z(a11)} &\textcolor{white}{\Leftrightarrow} &\textcolor{white}{P_Z(a11)} &\textcolor{white}{\Leftrightarrow} & \textcolor{white}{P_Z(a11)} & \textcolor{white}{\Leftrightarrow} & \textcolor{white}{P_Z(a11)} \\

O  & & \textcolor{green}{a1} & & O  & & \textcolor{green}{a2}  & & O & \Leftrightarrow & \textcolor{green}{a3}  & & O  \\
  & & \uparrow  & &   & & \uparrow   & &   & &   & &   \\
\textcolor{green}{a10} & & \textcolor{red}{d1} & & O & & \textcolor{red}{d2} & & O & & \textcolor{red}{d3} & \leftarrow & \textcolor{green}{a16} \\ 
\Updownarrow  & &   & &   & &    & &   & &   & &   \\
O & & \textcolor{green}{a4} &  & \textcolor{green}{a13} & &\textcolor{green}{a5} & & O & \Leftrightarrow & \textcolor{green}{a6} & & O  \\
  & & \uparrow  & & \Updownarrow  & & \uparrow   & &   & &   & &    \\
\textcolor{green}{a11} & & \textcolor{red}{d4} & & O & & \textcolor{red}{d5} & & O & & \textcolor{red}{d6} & \leftarrow & \textcolor{green}{a17} \\
  & &   & &   & &    & &   & &   & &     \\
O & & O & & O & \Leftrightarrow & \textcolor{green}{a14} & & O & & O & & O \\
  & &   & &   & &    & &   & &   & & \Updownarrow   \\
\textcolor{green}{a12} & \rightarrow & \textcolor{red}{d7} & & \textcolor{blue}{M_X(a15)} & & \textcolor{red}{d8} & & O & & \textcolor{red}{d9} & & \textcolor{green}{a18} \\
   & &   & &   & & \downarrow   & &   & & \downarrow  & &    \\
O & & \textcolor{green}{a7}& \Leftrightarrow & O & & \textcolor{green}{a8} & & O & & \textcolor{green}{a9} & & O \\
   & &   & &   & &    & &   & &   & &  

\end{array}
$

}

\vbox{

\small

Time step: 5
\\

$
\begin{array}{ccccccccccccc}
\textcolor{white}{P_Z(a11)}& \textcolor{white}{\Leftrightarrow} &\textcolor{white}{P_Z(a11)} & \textcolor{white}{\Leftrightarrow} &\textcolor{white}{P_Z(a11)} & \textcolor{white}{\Leftrightarrow} & \textcolor{white}{P_Z(a11)} &\textcolor{white}{\Leftrightarrow} &\textcolor{white}{P_Z(a11)} &\textcolor{white}{\Leftrightarrow} & \textcolor{white}{P_Z(a11)} & \textcolor{white}{\Leftrightarrow} & \textcolor{white}{P_Z(a11)} \\

O  & & \textcolor{blue}{M_Z(a1)} & & O  & & \textcolor{blue}{M_Z(a2)}  & & O & & \textcolor{green}{a3}  & & O  \\
  & &   & &   & &    & &   & & \uparrow  & &   \\
\textcolor{green}{a10} & \rightarrow & \textcolor{red}{d1} & & \textcolor{green}{a13} & & \textcolor{red}{d2} & & O & & \textcolor{red}{d3} & & \textcolor{blue}{M_X(a16)} \\ 
  & &   & & \Updownarrow  & &    & &   & &   & &   \\
O & & \textcolor{blue}{M_Z(a4)} &  & O & &\textcolor{blue}{M_Z(a5)} & & O & & \textcolor{green}{a6} & & O  \\
  & &  & &   & &    & &   & & \uparrow  & &    \\
\textcolor{green}{a11} & \rightarrow & \textcolor{red}{d4} & & O & & \textcolor{red}{d5} & & O & & \textcolor{red}{d6} & & \textcolor{blue}{M_X(a17)} \\
  & &   & &   & &  \uparrow  & &   & &   & &     \\
O & & O & & O & & \textcolor{green}{a14} & & O & & O & & O \\
  & &   & &   & &    & &   & &   & &    \\
\textcolor{blue}{M_X(a12)} & & \textcolor{red}{d7} & & O &  & \textcolor{red}{d8} & & O & & \textcolor{red}{d9} & \leftarrow & \textcolor{green}{a18} \\
   & & \downarrow  & &   & &    & &   & &   & &    \\
O & & \textcolor{green}{a7}&  & O & & \textcolor{blue}{M_Z(a8)} & & O & & \textcolor{blue}{M_Z(a9)} & & O \\
   & &   & &   & &    & &   & &   & &  

\end{array}
$

}

\vbox{

\small

Time step: 6
\\

$
\begin{array}{ccccccccccccc}
\textcolor{white}{P_Z(a11)}& \textcolor{white}{\Leftrightarrow} &\textcolor{white}{P_Z(a11)} & \textcolor{white}{\Leftrightarrow} &\textcolor{white}{P_Z(a11)} & \textcolor{white}{\Leftrightarrow} & \textcolor{white}{P_Z(a11)} &\textcolor{white}{\Leftrightarrow} &\textcolor{white}{P_Z(a11)} &\textcolor{white}{\Leftrightarrow} & \textcolor{white}{P_Z(a11)} & \textcolor{white}{\Leftrightarrow} & \textcolor{white}{P_Z(a11)} \\

O  & & O & & O  & & O  & & O & & \textcolor{blue}{M_Z(a3)}  & & O  \\
  & &   & &   & &    & &   & &  & &   \\
\textcolor{blue}{M_X(a10)} & & \textcolor{red}{d1} & & \textcolor{green}{a13} & \rightarrow  & \textcolor{red}{d2} & & O & & \textcolor{red}{d3} & & O \\ 
  & &   & &  & &    & &   & &   & &   \\
O & & O &  & O & & O & & O & & \textcolor{blue}{M_Z(a6)} & & O  \\
  & &  & &   & &    & &   & &   & &    \\
\textcolor{blue}{M_X(a11)} & & \textcolor{red}{d4} & & O & & \textcolor{red}{d5} & & O & & \textcolor{red}{d6} & & O \\
  & &   & &   & &    & &   & &   & &     \\
O & & O & & O & & \textcolor{blue}{M_X(a14)} & & O & & O & & O \\
  & &   & &   & &    & &   & &   & &    \\
O & & \textcolor{red}{d7} & &  & & \textcolor{red}{d8} & & O & & \textcolor{red}{d9} & & \textcolor{blue}{M_X(a18)} \\
   & &  & &   & &    & &   & &   & &    \\
O & & \textcolor{blue}{M_Z(a7)}&  & O & & O & & O & & O & & O \\
   & &   & &   & &    & &   & &   & &  

\end{array}
$

}

\vbox{

\small

Time step: 7
\\

$
\begin{array}{ccccccccccccc}
\textcolor{white}{P_Z(a11)}& \textcolor{white}{\Leftrightarrow} &\textcolor{white}{P_Z(a11)} & \textcolor{white}{\Leftrightarrow} &\textcolor{white}{P_Z(a11)} & \textcolor{white}{\Leftrightarrow} & \textcolor{white}{P_Z(a11)} &\textcolor{white}{\Leftrightarrow} &\textcolor{white}{P_Z(a11)} &\textcolor{white}{\Leftrightarrow} & \textcolor{white}{P_Z(a11)} & \textcolor{white}{\Leftrightarrow} & \textcolor{white}{P_Z(a11)} \\

O  & & O & & O  & & O  & & O & & O  & & O  \\
  & &   & &   & &    & &   & &  & &   \\
O & & \textcolor{red}{d1} & & \textcolor{blue}{M_X(a13)} & & \textcolor{red}{d2} & & O & & \textcolor{red}{d3} & & O \\ 
  & &   & &  & &    & &   & &   & &   \\
O & & O &  & O & & O & & O & & O & & O  \\
  & &  & &   & &    & &   & &   & &    \\
O & & \textcolor{red}{d4} & & O & & \textcolor{red}{d5} & & O & & \textcolor{red}{d6} & & O \\
  & &   & &   & &    & &   & &   & &     \\
O & & O & & O & & O & & O & & O & & O \\
  & &   & &   & &    & &   & &   & &    \\
O & & \textcolor{red}{d7} & & O & & \textcolor{red}{d8} & & O & & \textcolor{red}{d9} & & O \\
   & &  & &   & &    & &   & &   & &    \\
O & & O &  & O & & O & & O & & O & & O \\
   & &   & &   & &    & &   & &   & &  

\end{array}
$

}

\subsection{Preparation of $|\bar{0}\rangle$}

\vbox{

\small

Time step: 1
\\

$
\begin{array}{ccccccccccccc}
\textcolor{white}{P_Z(a11)}& \textcolor{white}{\Leftrightarrow} &\textcolor{white}{P_Z(a11)} & \textcolor{white}{\Leftrightarrow} &\textcolor{white}{P_Z(a11)} & \textcolor{white}{\Leftrightarrow} & \textcolor{white}{P_Z(a11)} &\textcolor{white}{\Leftrightarrow} &\textcolor{white}{P_Z(a11)} &\textcolor{white}{\Leftrightarrow} & \textcolor{white}{P_Z(a11)} & \textcolor{white}{\Leftrightarrow} & \textcolor{white}{P_Z(a11)} \\

O & & O &  & O & & O & & O & & O & & O \\
  & &   & &   & &    & &   & &  & &   \\
O  & & \textcolor{blue}{P_X(a4)} & & \textcolor{blue}{P_X(a13)}  & & \textcolor{blue}{P_X(a5)}  & & \textcolor{blue}{P_X(a14)} & & \textcolor{blue}{P_X(a6)}  & & \textcolor{blue}{P_X(a15)}  \\
  & &   & &   & &    & &   & &  & &   \\
O  & & \textcolor{blue}{P_Z(a7)} & & \textcolor{blue}{P_Z(a16)}  & & \textcolor{blue}{P_Z(a8)}  & & \textcolor{blue}{P_Z(a17)} & & \textcolor{blue}{P_Z(a9)}  & & \textcolor{blue}{P_Z(a18)}  \\

  & &   & &   & &    & &   & &  & &   \\
O  & & \textcolor{blue}{P_Z(d1)} & & \textcolor{blue}{P_Z(a19)}  & & \textcolor{blue}{P_Z(d2)}  & & \textcolor{blue}{P_Z(a20)} & & \textcolor{blue}{P_Z(d3)}  & & \textcolor{blue}{P_Z(a21)}  \\
  & &   & &   & &    & &   & &  & &   \\
O  & & \textcolor{blue}{P_X(d4)} & & \textcolor{blue}{P_X(a22)}  & & \textcolor{blue}{P_X(d5)}  & & \textcolor{blue}{P_X(a23)} & & \textcolor{blue}{P_X(d6)}  & & \textcolor{blue}{P_X(a24)}  \\
  & &   & &   & &    & &   & &  & &   \\
O & & O &  & O & & O & & O & & O & & O  \\
  & &   & &   & &    & &   & &  & &   \\

O & & O &  & O & & O & & O & & O & & O \\

\end{array}
$
\\
}
\vspace{0.2in}
\vbox{

\small

Time step: 2
\\

$
\begin{array}{ccccccccccccc}
\textcolor{white}{P_Z(a11)}& \textcolor{white}{\Leftrightarrow} &\textcolor{white}{P_Z(a11)} & \textcolor{white}{\Leftrightarrow} &\textcolor{white}{P_Z(a11)} & \textcolor{white}{\Leftrightarrow} & \textcolor{white}{P_Z(a11)} &\textcolor{white}{\Leftrightarrow} &\textcolor{white}{P_Z(a11)} &\textcolor{white}{\Leftrightarrow} & \textcolor{white}{P_Z(a11)} & \textcolor{white}{\Leftrightarrow} & \textcolor{white}{P_Z(a11)} \\

O  & & \textcolor{blue}{P_Z(a1)} & & \textcolor{blue}{P_Z(a10)}  & & \textcolor{blue}{P_Z(a2)}  & & \textcolor{blue}{P_Z(a11)} & & \textcolor{blue}{P_Z(a3)}  & & \textcolor{blue}{P_Z(a12)}  \\
  & &   & &   & &    & &   & &  & &   \\
O  & & \textcolor{green}{a4} & & \textcolor{green}{a13}  & & \textcolor{green}{a5}  & & \textcolor{green}{a14} & & \textcolor{green}{a6}  & & \textcolor{green}{a15}  \\
  & & \downarrow  & & \downarrow  & & \downarrow   & & \downarrow  & & \downarrow & & \downarrow  \\
O  & & \textcolor{green}{a7} & & \textcolor{green}{a16}  & & \textcolor{green}{a8}  & & \textcolor{green}{a17} & & \textcolor{green}{a9}  & & \textcolor{green}{a18}  \\

  & &   & &   & &    & &   & &  & &   \\
O  & & \textcolor{red}{d1} & & \textcolor{green}{a19}  & & \textcolor{red}{d2}  & & \textcolor{green}{a20} & & \textcolor{red}{d3}  & & \textcolor{green}{a21}  \\
  & & \uparrow   & & \uparrow  & & \uparrow   & & \uparrow  & & \uparrow & & \uparrow  \\
O  & & \textcolor{red}{d4} & & \textcolor{green}{a22}  & & \textcolor{red}{d5}  & & \textcolor{green}{a23} & & \textcolor{red}{d6}  & & \textcolor{green}{a24}  \\
  & &   & &   & &    & &   & &  & &   \\
O  & & \textcolor{blue}{P_Z(d7)} & & \textcolor{blue}{P_Z(a19)}  & & \textcolor{blue}{P_Z(d8)}  & & \textcolor{blue}{P_Z(a20)} & & \textcolor{blue}{P_Z(d9)}  & & \textcolor{blue}{P_Z(a21)}  \\
 & &   & &   & &    & &   & &  & &   \\

O & & O &  & O & & O & & O & & O & & O \\

\end{array}
$
\\

}
\vspace{0.2in}
\vbox{

\small

Time step: 3
\\

$
\begin{array}{ccccccccccccc}
\textcolor{white}{P_Z(a11)}& \textcolor{white}{\Leftrightarrow} &\textcolor{white}{P_Z(a11)} & \textcolor{white}{\Leftrightarrow} &\textcolor{white}{P_Z(a11)} & \textcolor{white}{\Leftrightarrow} & \textcolor{white}{P_Z(a11)} &\textcolor{white}{\Leftrightarrow} &\textcolor{white}{P_Z(a11)} &\textcolor{white}{\Leftrightarrow} & \textcolor{white}{P_Z(a11)} & \textcolor{white}{\Leftrightarrow} & \textcolor{white}{P_Z(a11)} \\

O  & & \textcolor{green}{a1} & & \textcolor{green}{a10}  & & \textcolor{green}{a2}  & & \textcolor{green}{a11} & & \textcolor{green}{a3}  & & \textcolor{green}{a12}  \\
  & & \uparrow   & & \uparrow  & & \uparrow   & & \uparrow  & & \uparrow & & \uparrow  \\
O  & & \textcolor{green}{a4} & & \textcolor{green}{a13}  & & \textcolor{green}{a5}  & & \textcolor{green}{a14} & & \textcolor{green}{a6}  & & \textcolor{green}{a15}  \\
  & &   & &   & &    & &   & &  & &   \\ 
O  & & \textcolor{green}{a7} & \leftarrow & \textcolor{green}{a16}  & & \textcolor{green}{a8}  & \leftarrow & \textcolor{green}{a17} & & \textcolor{green}{a9}  & \leftarrow & \textcolor{green}{a18}  \\

  & &   & &   & &    & &   & &  & &   \\
O  & & \textcolor{red}{d1} & \rightarrow & \textcolor{green}{a19}  & & \textcolor{red}{d2}  & \rightarrow & \textcolor{green}{a20} & & \textcolor{red}{d3}  & \rightarrow & \textcolor{green}{a21}  \\
 & &   & &   & &    & &   & &  & &   \\
O  & & \textcolor{red}{d4} & & \textcolor{green}{a22}  & & \textcolor{red}{d5}  & & \textcolor{green}{a23} & & \textcolor{red}{d6}  & & \textcolor{green}{a24}  \\
  & & \downarrow  & & \downarrow  & & \downarrow   & & \downarrow  & & \downarrow & & \downarrow  \\
O  & & \textcolor{red}{d7} & & \textcolor{green}{a19}  & & \textcolor{red}{d8}  & & \textcolor{green}{a20} & & \textcolor{red}{d9}  & & \textcolor{green}{a21}  \\
 & &   & &   & &    & &   & &  & &   \\

O & & O &  & O & & O & & O & & O & & O \\

\end{array}
$
\\
}
\vspace{0.2in}
\vbox{

\small

Time step: 4
\\

$
\begin{array}{ccccccccccccc}
\textcolor{white}{P_Z(a11)}& \textcolor{white}{\Leftrightarrow} &\textcolor{white}{P_Z(a11)} & \textcolor{white}{\Leftrightarrow} &\textcolor{white}{P_Z(a11)} & \textcolor{white}{\Leftrightarrow} & \textcolor{white}{P_Z(a11)} &\textcolor{white}{\Leftrightarrow} &\textcolor{white}{P_Z(a11)} &\textcolor{white}{\Leftrightarrow} & \textcolor{white}{P_Z(a11)} & \textcolor{white}{\Leftrightarrow} & \textcolor{white}{P_Z(a11)} \\

O  & & \textcolor{green}{a1} & \leftarrow & \textcolor{green}{a10}  & & \textcolor{green}{a2}  & \leftarrow & \textcolor{green}{a11} & & \textcolor{green}{a3}  & \leftarrow & \textcolor{green}{a12}  \\
  & &   & &   & &    & &   & &  & &   \\
O  & & \textcolor{green}{a4} & \leftarrow & \textcolor{green}{a13}  & & \textcolor{green}{a5}  & \leftarrow & \textcolor{green}{a14} & & \textcolor{green}{a6}  & \leftarrow & \textcolor{green}{a15}  \\
  & &   & &   & &    & &   & &  & &   \\ 
O  & & \textcolor{blue}{M_Z(a7)} & & \textcolor{green}{a16}  & & \textcolor{blue}{M_Z(a8)}  & & \textcolor{green}{a17} & & \textcolor{blue}{M_Z(a9)}  & & \textcolor{green}{a18}  \\

  & &   & &   & &    & &   & &  & &   \\
O  & & \textcolor{red}{d1} & & \textcolor{blue}{M_Z(a19)}  & & \textcolor{red}{d2}  &  & \textcolor{blue}{M_Z(a20)} & & \textcolor{red}{d3}  & & \textcolor{blue}{M_Z(a21)}  \\
 & &   & &   & &    & &   & &  & &   \\
O  & & \textcolor{red}{d4} & \rightarrow & \textcolor{green}{a22}  & & \textcolor{red}{d5}  & \rightarrow & \textcolor{green}{a23} & & \textcolor{red}{d6}  & \rightarrow & \textcolor{green}{a24}  \\
 & &   & &   & &    & &   & &  & &   \\
O  & & \textcolor{red}{d7} & \rightarrow & \textcolor{green}{a19}  & & \textcolor{red}{d8}  & \rightarrow & \textcolor{green}{a20} & & \textcolor{red}{d9}  & \rightarrow & \textcolor{green}{a21}  \\
 & &   & &   & &    & &   & &  & &   \\

O & & O &  & O & & O & & O & & O & & O \\

\end{array}
$
\\

}
\vspace{0.2in}
\vbox{

\small

Time step: 5
\\

$
\begin{array}{ccccccccccccc}
\textcolor{white}{P_Z(a11)}& \textcolor{white}{\Leftrightarrow} &\textcolor{white}{P_Z(a11)} & \textcolor{white}{\Leftrightarrow} &\textcolor{white}{P_Z(a11)} & \textcolor{white}{\Leftrightarrow} & \textcolor{white}{P_Z(a11)} &\textcolor{white}{\Leftrightarrow} &\textcolor{white}{P_Z(a11)} &\textcolor{white}{\Leftrightarrow} & \textcolor{white}{P_Z(a11)} & \textcolor{white}{\Leftrightarrow} & \textcolor{white}{P_Z(a11)} \\

O  & & \textcolor{blue}{M_Z(a1)} &  & \textcolor{green}{a10}  & & \textcolor{blue}{M_Z(a2)}  &  & \textcolor{green}{a11} & & \textcolor{blue}{M_Z(a3)}  & & \textcolor{green}{a12}  \\
  & &   & &   & &    & &   & &  & &   \\
O  & & \textcolor{blue}{M_Z(a4)} &  & \textcolor{green}{a13}  & & \textcolor{blue}{M_Z(a5)}  & & \textcolor{green}{a14} & & \textcolor{blue}{M_Z(a6)}  & & \textcolor{green}{a15}  \\
  & &   & &   & &    & &   & &  & &   \\ 
O  & & \textcolor{red}{d1} & & O  & & \textcolor{red}{d2}  & & O & & \textcolor{red}{d3}  & & O  \\

  & & \Updownarrow  & &  \Updownarrow  & &  \Updownarrow   & &  \Updownarrow  & &  \Updownarrow & &  \Updownarrow  \\
O  & & O & & \textcolor{green}{a16}  & & O  &  &  \textcolor{green}{a17} & & O  & &  \textcolor{green}{a18}  \\
 & &   & &   & &    & &   & &  & &   \\
O  & & \textcolor{red}{d4} &  & \textcolor{blue}{M_Z(a22)}  & & \textcolor{red}{d5}  &  & \textcolor{blue}{M_Z(a23)} & & \textcolor{red}{d6}  &  & \textcolor{blue}{M_Z(a24)}  \\
 & &   & &   & &    & &   & &  & &   \\
O  & & \textcolor{red}{d7} &  & \textcolor{blue}{M_Z(a19)}  & & \textcolor{red}{d8}  &  & \textcolor{blue}{M_Z(a20)} & & \textcolor{red}{d9}  &  & \textcolor{blue}{M_Z(a21)}  \\
 & &   & &   & &    & &   & &  & &   \\

O & & O &  & O & & O & & O & & O & & O \\

\end{array}
$
\\

}
\vspace{0.2in}
\vbox{

\small

Time step: 6
\\

$
\begin{array}{ccccccccccccc}
\textcolor{white}{P_Z(a11)}& \textcolor{white}{\Leftrightarrow} &\textcolor{white}{P_Z(a11)} & \textcolor{white}{\Leftrightarrow} &\textcolor{white}{P_Z(a11)} & \textcolor{white}{\Leftrightarrow} & \textcolor{white}{P_Z(a11)} &\textcolor{white}{\Leftrightarrow} &\textcolor{white}{P_Z(a11)} &\textcolor{white}{\Leftrightarrow} & \textcolor{white}{P_Z(a11)} & \textcolor{white}{\Leftrightarrow} & \textcolor{white}{P_Z(a11)} \\

O  & & O &  & \textcolor{green}{a10}  & & O  &  & \textcolor{green}{a11} & & O  & & \textcolor{green}{a12}  \\
  & &   & &   & &    & &   & &  & &   \\
O  & & \textcolor{red}{d1} &  & O  & & \textcolor{red}{d2}  & & O & & \textcolor{red}{d3}  & & O  \\
  & & \Updownarrow  & &  \Updownarrow  & &  \Updownarrow   & &  \Updownarrow  & &  \Updownarrow & &  \Updownarrow  \\

O  & & O & & \textcolor{green}{a13} & & O  & & \textcolor{green}{a14} & & O  & & \textcolor{green}{a15}  \\
  & &   & &   & &    & &   & &  & &   \\

O  & & \textcolor{red}{d4} & & O  & & \textcolor{red}{d5}  &  &  O & & \textcolor{red}{d6}  & &  O  \\
  & & \Updownarrow  & &  \Updownarrow  & &  \Updownarrow   & &  \Updownarrow  & &  \Updownarrow & &  \Updownarrow  \\
O  & & O &  & \textcolor{green}{a16}  & & O  &  & \textcolor{green}{a17} & & O  &  & \textcolor{green}{a18}  \\
 & &   & &   & &    & &   & &  & &   \\
O  & & \textcolor{red}{d7} &  & O  & & \textcolor{red}{d8}  &  & O & & \textcolor{red}{d9}  &  & O  \\
 & &   & &   & &    & &   & &  & &   \\

O & & O &  & O & & O & & O & & O & & O \\

\end{array}
$
\\

}
\vspace{0.2in}
\vbox{

\small

Time step: 7
\\

$
\begin{array}{ccccccccccccc}
\textcolor{white}{P_Z(a11)}& \textcolor{white}{\Leftrightarrow} &\textcolor{white}{P_Z(a11)} & \textcolor{white}{\Leftrightarrow} &\textcolor{white}{P_Z(a11)} & \textcolor{white}{\Leftrightarrow} & \textcolor{white}{P_Z(a11)} &\textcolor{white}{\Leftrightarrow} &\textcolor{white}{P_Z(a11)} &\textcolor{white}{\Leftrightarrow} & \textcolor{white}{P_Z(a11)} & \textcolor{white}{\Leftrightarrow} & \textcolor{white}{P_Z(a11)} \\

O  & & O &  & O  & & O  &  & O & & O  & & O  \\
  & &   & & \Updownarrow  & &    & & \Updownarrow  & &  & & \Updownarrow   \\
O  & & \textcolor{red}{d1} &  & \textcolor{green}{a10}  & & \textcolor{red}{d2}  & & \textcolor{green}{a11} & & \textcolor{red}{d3}  & & \textcolor{green}{a12}  \\

  & &   & &   & &    & &   & &  & &   \\ 
O  & & O & & O & & O  & & O & & O  & & O  \\
  & &   & & \Updownarrow  & &    & & \Updownarrow  & &  & & \Updownarrow   \\

O  & & \textcolor{red}{d4} & & \textcolor{green}{a13}  & & \textcolor{red}{d5}  &  &  \textcolor{green}{a14} & & \textcolor{red}{d6}  & &  \textcolor{green}{a15}  \\
 & &   & &   & &    & &   & &  & &   \\
O  & & O &  & O  & & O  &  & O & & O  &  & O  \\
  & &   & & \Updownarrow  & &    & & \Updownarrow  & &  & & \Updownarrow   \\

O  & & \textcolor{red}{d7} &  & \textcolor{green}{a16}  & & \textcolor{red}{d8}  &  & \textcolor{green}{a17} & & \textcolor{red}{d9}  &  &  \textcolor{green}{a18} \\
 & &   & &   & &    & &   & &  & &   \\

O & & O &  & O & & O & & O & & O & & O \\

\end{array}
$
\\

}
\vspace{0.2in}
\vbox{

\small

Time step: 8
\\

$
\begin{array}{ccccccccccccc}
\textcolor{white}{P_Z(a11)}& \textcolor{white}{\Leftrightarrow} &\textcolor{white}{P_Z(a11)} & \textcolor{white}{\Leftrightarrow} &\textcolor{white}{P_Z(a11)} & \textcolor{white}{\Leftrightarrow} & \textcolor{white}{P_Z(a11)} &\textcolor{white}{\Leftrightarrow} &\textcolor{white}{P_Z(a11)} &\textcolor{white}{\Leftrightarrow} & \textcolor{white}{P_Z(a11)} & \textcolor{white}{\Leftrightarrow} & \textcolor{white}{P_Z(a11)} \\

O  & & O &  & O  & & O  &  & O & & O  & & O  \\
  & &   & & & &    & &  & &  & &   \\
O  & & \textcolor{red}{d1} & \leftarrow  & \textcolor{green}{a10}  & & \textcolor{red}{d2}  & \leftarrow & \textcolor{green}{a11} &  & \textcolor{red}{d3}  & \leftarrow & \textcolor{green}{a12}  \\

  & &   & &   & &    & &   & &  & &   \\ 
O  & & O & & O & & O  & & O & & O  & & O  \\
  & &   & &   & &    & &  & &  & &    \\

O  & & \textcolor{red}{d4} & \leftarrow & \textcolor{green}{a13}  & & \textcolor{red}{d5}  & \leftarrow &  \textcolor{green}{a14} & & \textcolor{red}{d6}  & \leftarrow &  \textcolor{green}{a15}  \\
 & &   & &   & &    & &   & &  & &   \\
O  & & O &  & O  & & O  &  & O & & O  &  & O  \\
  & &   & &   & &    & &   & &  & &   \\

O  & & \textcolor{red}{d7} &  \leftarrow & \textcolor{green}{a16}  & & \textcolor{red}{d8}  & \leftarrow & \textcolor{green}{a17} & & \textcolor{red}{d9}  & \leftarrow &  \textcolor{green}{a18} \\
 & &   & &   & &    & &   & &  & &   \\

O & & O &  & O & & O & & O & & O & & O \\

\end{array}
$
\\

}
\vspace{0.2in}
\vbox{

\small

Time step: 9
\\

$
\begin{array}{ccccccccccccc}
\textcolor{white}{P_Z(a11)}& \textcolor{white}{\Leftrightarrow} &\textcolor{white}{P_Z(a11)} & \textcolor{white}{\Leftrightarrow} &\textcolor{white}{P_Z(a11)} & \textcolor{white}{\Leftrightarrow} & \textcolor{white}{P_Z(a11)} &\textcolor{white}{\Leftrightarrow} &\textcolor{white}{P_Z(a11)} &\textcolor{white}{\Leftrightarrow} & \textcolor{white}{P_Z(a11)} & \textcolor{white}{\Leftrightarrow} & \textcolor{white}{P_Z(a11)} \\

O  & & O &  & O  & & O  &  & O & & O  & & O  \\
  & &   & & & &    & &  & &  & &   \\
O  & & \textcolor{red}{d1} &  & \textcolor{blue}{M_X(a10)}  & & \textcolor{red}{d2}  & & \textcolor{blue}{M_X(a11)} &  & \textcolor{red}{d3}  & & \textcolor{blue}{M_X(a12)}  \\

  & &   & &   & &    & &   & &  & &   \\ 
O  & & O & & O & & O  & & O & & O  & & O  \\
  & &   & &   & &    & &  & &  & &    \\

O  & & \textcolor{red}{d4} &  & \textcolor{blue}{M_X(a13)}  & & \textcolor{red}{d5}  & &  \textcolor{blue}{M_X(a14)} & & \textcolor{red}{d6}  &  &  \textcolor{blue}{M_X(a15)}  \\
 & &   & &   & &    & &   & &  & &   \\
O  & & O &  & O  & & O  &  & O & & O  &  & O  \\
  & &   & &   & &    & &   & &  & &   \\

O  & & \textcolor{red}{d7} &  & \textcolor{blue}{M_X(a16)}  & & \textcolor{red}{d8}  &  & \textcolor{blue}{M_X(a17)} & & \textcolor{red}{d9}  &  &  \textcolor{blue}{M_X(a18)} \\
 & &   & &   & &    & &   & &  & &   \\

O & & O &  & O & & O & & O & & O & & O \\

\end{array}
$
\\
}

\bibliographystyle{prsty}
\bibliography{latency}

\begin{thebibliography}{10}

\bibitem{svore2007a}
K.~M. Svore, D.~P. Di{V}incenzo, and B.~M. Terhal, Quantum Information \&
  Computation {\bf 7},  297  (2007).

\bibitem{nielsen2000}
M.~N. Nielsen and I.~L. Chuang, {\em Quantum computation and quantum
  information} (Cambridge University Press, Cambridge, 2000).

\bibitem{preskill1998a}
J. Preskill, Proc. R. Soc. Lond. A {\bf 454},  385  (1998),
  {\tt{quant-ph/9705031}}.

\bibitem{aharonov1997a}
D. Aharonov and M. Ben-Or,  in {\em STOC '97: Proceedings of the twenty-ninth
  annual ACM symposium on Theory of computing} (ACM, New York, NY, USA, 1997),
  pp.\ 176--188.

\bibitem{steane2003a}
A.~M. Steane, Phys. Rev. A {\bf 68},  042322  (2003).

\bibitem{gottesman1998a}
D. Gottesman, Phys. Rev. A {\bf 57},  127  (1998).

\bibitem{szkopek2006a}
T. Szkopek {\it et~al.}, Nanotechnology, IEEE Transactions on {\bf 5},  42
  (Jan. 2006).

\bibitem{shor1995a}
P.~W. Shor, Phys. Rev. A {\bf 52},  R2493  (1995).

\bibitem{bacon2006a}
D. Bacon, Physical Review A {\bf 73},  012340  (2006).

\bibitem{aliferis2007a}
P. Aliferis and A.~W. Cross, Physical Review Letters {\bf 98},  220502  (2007).

\bibitem{knill2005a}
E. Knill, Nature {\bf 434},  39  (2005).

\bibitem{aliferis2006a}
P. Aliferis, D. Gottesman, and J. Preskill, Quantum Information \& Computation
  {\bf 6},  97  (2006).

\end{thebibliography}

\end{document}